 \documentclass[final,3p,times,twocolumn]{elsarticle}

\usepackage{epsfig}

\usepackage{amssymb}
 \usepackage{amsthm}
 \usepackage{amsmath}
\usepackage{xcolor}  

\journal{     }

\begin{document}

\begin{frontmatter}



\title{PHOTONEUTRON REACTIONS $^{181}\rm{Ta}(\gamma,\textit{x}n; \textit{x} = 1 \div 8)^{181-\textit{x}}\rm{Ta}$ AT  \mbox{$E_{\rm{\gamma max}}$ = 80 $\div$ 95 MeV}}


\author{A.N. Vodin,O.S. Deiev }

\author{I.S.~Timchenko\corref{cor1}}
\ead{timchenko@kipt.kharkov.ua}
 \cortext[cor1]{Corresponding author}

\author{S.N. Olejnik, \\ M.I. Ayzatskiy, V.A. Kushnir, V.V. Mitrochenko, S.A. Perezhogin}

\address{National Science Center "Kharkov Institute of Physics and Technology", \\
 1 Akademicheskaya St., 61108 Kharkov, Ukraine}

\begin{abstract}
The bremsstrahlung flux-averaged cross-sections $\langle{\sigma(E_{\rm{\gamma max}})}\rangle$ for the $^{181}\rm{Ta}(\gamma,\textit{x}n; \textit{x} = 1 \div 8)^{181-\textit{x}}\rm{Ta}$ photoneutron reactions have been measured at end-point bremsstrahlung energies ranging from 80~MeV to 95~MeV. The measurements were performed with the beam from the NSC KIPT electron linear accelerator LUE-40 using the residual $\gamma$-activity method. The theoretical  $\langle{\sigma(E_{\rm{\gamma max}})}\rangle$ values were computed using the cross-sections  $\sigma(E)$ from TALYS1.9 code. A comparison between the measured cross-sections  $\langle{\sigma(E_{\rm{\gamma max}})}\rangle$ and the theoretical values has demostrated their good agreement for the reactions with escape of up to 6 neutrons, and substantial differences for the $(\gamma,7\rm{n})$ and $(\gamma,8\rm{n})$  reactions. Isomeric ratios of the average cross-sections $d(E_{\rm{\gamma max}})$ have been found for the  $^{181}\rm{Ta}(\gamma,3n)^{178g,m}\rm{Ta}$ reaction products. The results have been compared with the literature data and the computations based on TALYS1.9 code.  
\end{abstract}




\end{frontmatter}

\section{Introduction}
\label{intro}

Most of the nuclear photodisintegration studies in a wide mass region have been dedicated to investigation of the properties and nature of the giant dipole resonance (GDR) with the use of bremsstrahlung/quasimonoenergetic photon beams \cite{1,2}. At the same time, it is of interest to investigate multiparticle photonucleon reactions on nuclei in the energy range above the GDR and up to the pion production threshold ($E_{\rm{th}} \approx 145$~MeV). This is linked with the change in the mechanism of photon interaction with nuclei in this energy region. Thus, fundamental information can be obtained about the competition between two mechanisms of nuclear photodisintegration, namely, through the GDR excitation and the quasideuteron photoabsorption. The observation of the mentioned reactions, which show relatively low cross-section values, calls for intensive beams of incident $\gamma$-quanta that can be provided by electron linear accelerators as the high-energy electrons pass through the target-converter.  Some results of investigations of multiparticle nuclear reactions can be found, for example, in \cite{3,4}.

The studies of photoneutron reactions on $^{181}\rm{Ta}$ are of interest, because this nucleus is strongly deformed, with the quadrupole deformation parameter being $\beta = 0.26$ \cite{5}. This strong deformation leads, for example, to a complex energy-dependence structure of the $^{181}\rm{Ta}(\gamma,n)^{180}\rm{Ta}$ reaction cross-section, which shows two maxima in the GDR region. Consequently, this may cause the change in the form of the energy dependences of the reaction yield or the bremsstrahlung flux-averaged cross-section. Besides, the  $^{180\rm{m}}\rm{Ta}$  isotope belongs to the so-called “by-passed” nuclei, the existence of which in nature poses the problems of searching and investigating the processes of heavy nuclei formation unrelated to neutron capture. The measurements of the  $(\gamma,\rm{n})$ reaction cross-section in the energy region above the GDR may give a new insight into the properties of the high-energy part of the GDR and improve the theoretical description of the nucleosynthesis rate responsible for the $p$-production of $^{180\rm{m}}\rm{Ta}$  -- the  rarest stable nucleus in nature. 
   
The experiments on photodisintegration of $^{181}\rm{Ta}$ in the GDR region were performed in a variety of works  \cite{6,7,8,9,10,11,12,13} with the use of bremsstrahlung/quasimonoenergetic photon beams. For example, the data on photoabsorption cross-sections  $\sigma(\gamma,\rm{abs})$ were obtained in ref.~\cite{6} by the absorption method; the measurements in Livermore \cite{7} and Saclay \cite{8} were carried out using quasimonochromatic photons; in refs.~\cite{9,10,11,12,13}, to obtain the $\sigma(\gamma,s\rm{n})$ values, bremsstrahlung photon beams were used. The cross-section $\sigma(\gamma,\rm{abs})$, evaluated as the sum of cross-sections for all the investigated photonucleon reactions  on $^{181}\rm{Ta}$, has been determined in ref.~\cite{14} from the analysis of photonucleon reactions of different multiplicity.

The partial cross-sections for  $(\gamma,\rm{n})$ and $ (\gamma,2\rm{n})$ photoneutron reactions on $^{181}\rm{Ta}$ were obtained in ref.~\cite{7} by the method of direct neutron registration in the GDR region. Besides, the studies on the partial cross-sections for the $^{181}\rm{Ta}(\gamma,n)$, $(\gamma,2\rm{n})$ and $(\gamma,3\rm{n})$ reactions, as well as for the $(\gamma,4\rm{n})$ reaction up to an energy of 36~MeV, were carried out in work \cite{8}. The cross-section data obtained in those two works (\cite{7} and \cite{8}) diverge considerably from each other. For example, for the case of $(\gamma,\rm{n})$ reaction the difference at the maximum  of the cross-section reaches $\sim\!25\%$. This divergence may be due to the peculiarity of the direct neutron registration method, in which it is difficult to separate neutrons from the $(\gamma,\rm{n})$ and $(\gamma,2\rm{n})$ reactions. 

The cross-sections for the photoneutron reactions on tantalum at energies above the GDR were investigated in some experiments using the bremsstrahlung beams. So, in ref.~\cite{15}, at the end-point bremsstrahlung energy $E_{\rm{\gamma max}} = 55$~MeV the bremsstrahlung flux-averaged yields were obtained for the reactions $^{181}\rm{Ta}(\gamma,3n)^{178g,m}\rm{Ta}$ and 
 $^{181}\rm{Ta}(\gamma,p)^{180m}\rm{Hf}$.  The data on relative yields were found at 
 $E_{\rm{\gamma max}} = 67.7$~MeV in ref.~\cite{16} for the $^{181}\rm{Ta}(\gamma,\textit{x}n)^{181-\textit{x}}\rm{Ta}$ reactions with escape of up to 6 neutrons, and also for the reactions with the charged particle in the exit channel: $^{181}\rm{Ta}(\gamma,p)^{180m}\rm{Hf}$ and $^{181}\rm{Ta}(\gamma,pn)^{179m}\rm{Hf}$. In those studies, the comparison was made between the experimental data and the theoretical computations with the codes TALYS1.7~\cite{17} and EMPIRE-3.2 \cite{18}, and also, with the use of the combined model \cite{19}. Note that the cross-sections for photoproton and photoneutron reactions on the  $^{181}\rm{Ta}$ nucleus with a larger multiplicity $(\textit{x} > 4)$ were not measured \cite{20}. The isomeric ratio values for the  $^{181}\rm{Ta}(\gamma,3n)^{178g,m}\rm{Ta}$ reaction products have been presented in papers \cite{15,16,21,22}.

At the same time, it should be pointed out that the experiments on bremsstrahlung beams considerably complicate the procedure of photonuclear reaction cross-section determination. First of all, it is necessary to carry out correct calculations of the $\gamma$-flux density that would correspond to actual experimental conditions, using the modern computational techniques (e.g., GEANT4). Aside from that, the said experiments involve the measurements of integral characteristics of the reactions, and this calls for additional mathematical treatment of the results. And yet, in spite of the encountered difficulties, the bremsstrahlung beams remain an important tool in photonuclear reaction studies. 

The photonuclear reaction data obtained with the use of bremsstrahlung beams by the residual $\gamma$-activity  method can be represented in the forms of the relative $Y(E_{\rm{\gamma max}})$ reaction yield \cite{16}, the bremsstrahlung flux-averaged cross-section $\langle{\sigma(E_{\rm{\gamma max}})}\rangle$ \cite{24,25} or the average cross-section per equivalent photon $\langle{\sigma(E_{\rm{\gamma max}})_{\rm{Q}}}\rangle$ \cite{26}. The analysis of the relative  $Y(E_{\rm{\gamma max}})$ data is restricted by their representation as dimensionless normalized yields, generally normalized to the $(\gamma,\rm{n})$ reaction yield. The cross-sections $\langle{\sigma(E_{\rm{\gamma max}})_{\rm{Q}}}\rangle$ are convenient to use in the case of the final nucleus-product formation in several reaction channels \cite{27}. For single-channel reactions, the representation of $\langle{\sigma(E_{\rm{\gamma max}})}\rangle$ permits a more detailed investigation of the cross-section as a function of the bremsstrahlung energy, since it is insensitive to the low-energy part of the bremsstrahlung spectrum.  

The present work is concerned with the measurements of bremsstrahlung flux-averaged cross-sections $\langle{\sigma(E_{\rm{\gamma max}})}\rangle$ for $^{181}\rm{Ta}$ photoneutron reactions with escape of up to 8 neutrons, and also, of isomeric ratios of average cross-sections $d(E_{\rm{\gamma max}})$ for the $^{181}\rm{Ta}(\gamma,3n)^{178g,m}\rm{Ta}$ reaction products at end-point energies of bremsstrahlung spectra  $E_{\rm{\gamma max}} = 80 \div 95$~MeV.

\section{Experimental procedure}
\label{sec:1}
\begin{figure}[h]
   	\resizebox{0.49\textwidth}{!}{%
  \includegraphics{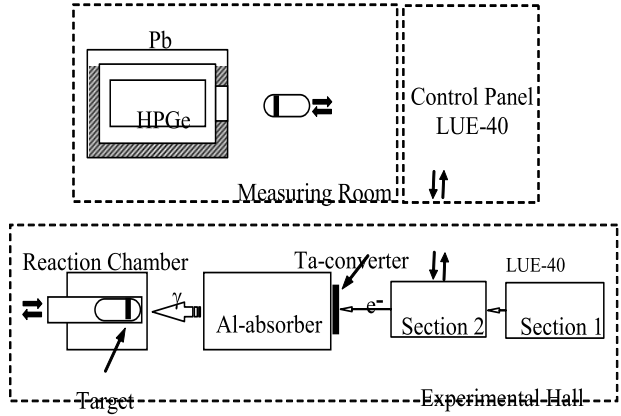}}
   	\caption{Schematic block diagram of the experiment. The upper part shows the measuring room and the room for accelerator performance control. The lower part shows (from right to left) two sections of the accelerator LUE-40; Ta-converter; Al-absorder; bombardment chamber.}
   	\label{fig1}
   \end{figure}

The experiments were performed on the bremsstrahlung beam from the NSC KIPT electron linear accelerator LUE-40 \cite{28,29}. The experimental layout is shown in Fig.~\ref{fig1}. Electrons of initial energy $E_{e^-}$ were incident on the target-converter, made from natural tantalum plate with cross dimensions of 20 mm×20 mm, and 1.05 mm in thickness. To remove the electrons from the bremsstrahlung flux, a cylinder-shaped Al-absorber, 100 mm in diameter and 150 mm in length, was used. 

  \begin{table*}[]
      	\caption{\label{tab1} Spectroscopic data \cite{31} on the nuclei-products from the reactions $^{181}\rm{Ta}(\gamma,\textit{x}n; \textit{x} = 1 \div 8)^{181-\textit{x}}\rm{Ta}$ and the monitoring reaction $^{100}\rm{Mo}(\gamma,n)^{99}\rm{Mo}$}
      	\centering
      	\begin{tabular}{cccccc}
      		\hline	\vspace{1ex}
      		 \begin{tabular}{c} Nuclear \\  reaction \end{tabular} & $E_{\rm{th}}$,~MeV & \begin{tabular}{c} $J^\pi$ of  \\ nucleus-product
\end{tabular} & $T_{1/2}$ & $E_{\gamma}$,~keV & $I_{\gamma}$, \% \\ 	\hline
&&&&&\\ 	   
$^{181}\rm{Ta}(\gamma,n)^{180g}\rm{Ta}$ & 7.58 & $1^+$ & 8.152 (6) h & 103.557 (7) & 0.81 (16) \\
$^{181}\rm{Ta}(\gamma,n)^{180m}\rm{Ta}$ & 7.65 & $9^-$ & $>1.2 \cdot 10^{15}$ y &   &   \\ 
$^{181}\rm{Ta}(\gamma,2n)^{179}\rm{Ta}$ & 14.22 & $7/2^+$ & 1.82 (3) y & 54.611  & 13.6 (4)   \\
&&&&  55.790 & 23.7 (7) \\
$^{181}\rm{Ta}(\gamma,3n)^{178g}\rm{Ta}$ & 22.21 & $1^+$ & 9.31 (3) min & 1350.68 (3) & 1.18 (3) \\
$^{181}\rm{Ta}(\gamma,3n)^{178m}\rm{Ta}$ & 22.24 & $(7)^-$ & 2.36 (8) h & 426.383 (6) & 97.0 (13) \\
$^{181}\rm{Ta}(\gamma,4n)^{177}\rm{Ta}$ & 29.01 & $ 7/2^+$ & 56.56 (6) h & 112.9498 (5) & 7.2 (8) \\
$^{181}\rm{Ta}(\gamma,5n)^{176}\rm{Ta}$ & 37.44 & $ (1)^-$ & 8.09 (5) h & 1159.28 (9) & 25 (?) \\
$^{181}\rm{Ta}(\gamma,6n)^{175}\rm{Ta}$ & 44.46 & $ 7/2^+$ & 10.5 (2) h & 348.5 (5) & 12.0 (6) \\
$^{181}\rm{Ta}(\gamma,7n)^{174}\rm{Ta}$ & 53.21 & $ 3^+$ & 1.05 (3) h & 1205.92 (4) & 4.9 (4) \\
$^{181}\rm{Ta}(\gamma,8n)^{173}\rm{Ta}$ & 60.63 & $ 5/2^-$ & 3.14 (13) h & 172.2 (1) & 18 (?) \\
$^{100}\rm{Mo}(\gamma,n)^{99}\rm{Mo}$   & 8.29   & $1/2^+$  & 65.94 (1) h & 739.50 (2) & 12.13 (12)  \\ 
	\hline
      	\end{tabular}	        
  \end{table*}

In the experiments, the method for measuring the residual $\gamma$-activity of the irradiated sample has been used, which provided a simultaneous data acquisition from different channels of photonuclear reactions. This method is well known and has been described in a variety of papers dedicated to the studies of multiparticle photonuclear reactions, e.g., on the $^{93}$Nb \cite{24,Nb}.  

The bremsstrahlung spectra were calculated by means of the open certified programme code GEANT4 \cite{30} with due regard to the real geometry of the experiment, where spatial and energy distributions of the electron beam were taken into account. In addition, the bremsstrahlung flux was estimated from the yield of the $^{100}\rm{Mo}(\gamma,n)^{99}\rm{Mo}$ reaction. For this purpose, the natural molybdenum target-witness, placed close by the target under study, was exposed to radiation.

The natural tantalum/molybdenum targets of diameter 8 mm were arranged behind the Al-absorber on the electron beam axis. For transporting the targets to the place of irradiation and back for induced activity registration the pneumatic tube transfer system was used. Experimentally, this enabled us to obtain the data on the $^{181}\rm{Ta}(\gamma,3n)^{178g}\rm{Ta}$ reaction with a relatively short half-life period (see Table~\ref{tab1}).

\begin{figure*}[]
  	\resizebox{1.02\textwidth}{!}{%
  \includegraphics{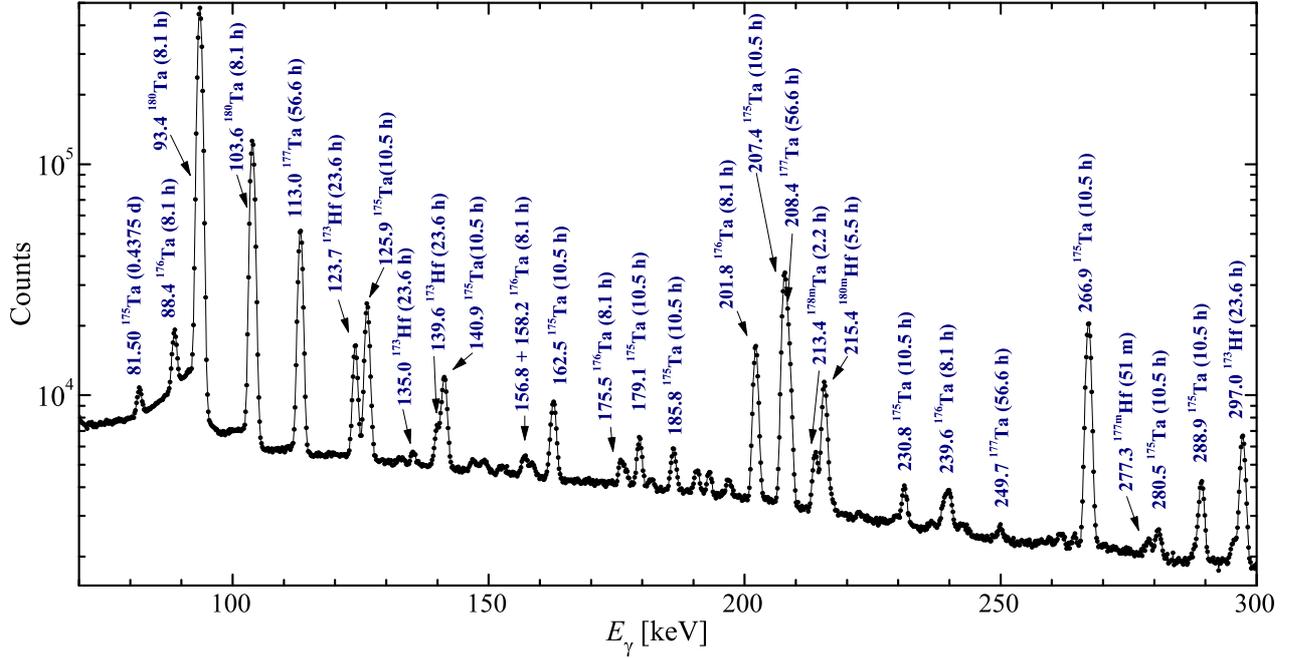}}
	\caption{Fragment of  $\gamma$-radiation spectrum from the tantalum target of mass 43.665~mg after exposure to the bremsstrahlung flux.  $E_{\rm{\gamma max}} = 85.6$~MeV, $t_{\rm{irr}} = 30$~min.}
	\label{fig2}

\end{figure*}

 In the experiments, six pairs of natural tantalum/ molybdenum samples were exposed to radiation at different end-point bremsstrahlung energies  $E_{\rm{\gamma max}}$  in the range from 80 to 95~MeV. The masses of Ta target and Mo target were, respectively, $\sim 43$~mg and $\sim 60$~mg. The time of irradiation  $t_{\rm{irr}}$ and the time of residual $\gamma$-activity spectrum measurement $t_{\rm{meas}}$ were both 30 min. To exemplify, Fig.~\ref{fig2} shows a fragment of the $\gamma$-spectrum from the tantalum target.

For $\gamma$-radiation registration, a semiconductor HPGe detector (Canberra GC-2018) was used with resolutions of  0.8 keV and 1.8 keV (FWHM) for the energies $E_{\gamma} =122$~keV and 1332~keV, respectively. The absolute registration efficiency of the detector was calibrated with a standard set of $\gamma$-ray sources $^{22}$Na, $^{60}$Co, $^{133}$Ba, $^{137}$Cs, $^{152}$Eu and $^{241}\!\rm{Am}$.

 To investigate the reactions of interest, the residual activity $\gamma$-spectrum of the irradiated target was analyzed, and the activities $\triangle A$ (i.e., the number of counts of $\gamma$-quanta in the full absorption peak) were determined for the $\gamma$-lines corresponding to the isotopes produced. Relying on the data from ref.~\cite{31}, Table~\ref{tab1} gives the parameters of both the reactions under study and the monitoring reaction, viz., the energy $E_{\gamma}$ and intensity $I_{\gamma}$ of the $\gamma$-lines in use.

Based on the obtained activities  $\triangle A$, we determined the experimental values of the bremsstrahlung flux-averaged cross-sections by the expression: 

\begin{equation}\begin{split}
\langle{\sigma(E_{\rm{\gamma max}})}\rangle = \\
\frac{\lambda \triangle A  {\rm{\Phi}}^{-1}(E_{\rm{\gamma max}})}{N_x I_{\gamma} \ \varepsilon (1-\exp(-\lambda t_{\rm{irr}}))\exp(-\lambda t_{\rm{cool}})(1-\exp(-\lambda t_{\rm{meas}}))},
\label{form1}
\end{split}
\end{equation}
where ${\rm{\Phi}}(E_{\rm{\gamma max}}) = {\int\limits_{E_{\rm{th}}}^{E_{\rm{\gamma max}}}W(E,E_{\rm{\gamma max}})dE}$  is the integrated bremsstrahlung flux $W(E,E_{\rm{\gamma max}})$ in the energy range from the reaction threshold $E_{\rm{th}}$ of the corresponding reaction up to the maximum energy of ${\gamma}$-quanta $E_{\rm{\gamma max}}$, $N_x$ is the number of target atoms, $I_{\gamma}$ is the absolute intensity of the analyzed $\gamma$-quanta, $\varepsilon$ is the absolute detection efficiency for the analyzed $\gamma$-quanta energy, $\lambda$ is the decay constant \mbox{($\rm{ln}2/\textit{T}_{1/2}$)}; $t_{\rm{irr}}$, $t_{\rm{cool}}$ and $t_{\rm{meas}}$ are the irradiation time, cooling time and measurement time, respectively.

The cross-sections $\langle{\sigma(E_{\rm{\gamma max}})}\rangle$ obtained in this way were compared with the theoretical values, which were based on the reaction cross-sections $\sigma(E)$ from the TALYS1.9 code with the default options: Constant temperature + Fermi gas model \cite{32}. For this purpose, the values of the $\sigma(E)$  were weighted over the bremsstrahlung flux   $W(E,E_{\rm{\gamma max}})$ from the reaction threshold $E_{\rm{th}}$ up to $E_{\rm{\gamma max}}$. As a result of this procedure, the theretical bremsstrahlung flux-averaged cross-sections $\langle{\sigma(E_{\rm{\gamma max}})}\rangle$  were obtained according to the formula:
 \begin{equation}\label{form2}
\langle{\sigma(E_{\rm{\gamma max}})}\rangle =   {\rm{\Phi}}^{-1}(E_{\rm{\gamma max}}) \int\limits_{E_{\rm{th}}}^{E_{\rm{\gamma max}}}\sigma(E)\cdot W(E,E_{\rm{\gamma max}})dE.
\end{equation}
A more detailed description of all the procedures necessary for determining experimental and theoretical values of $\langle{\sigma(E_{\rm{\gamma max}})}\rangle$ can be found in refs.~\cite{24,Nb,34}.

If the nucleus-product has an isomeric state, the total average cross-section $\langle{\sigma(E_{\rm{\gamma max}})}\rangle_{\rm{tot}}$ for the reaction under study (hereafter referred to as $\langle{\sigma(E_{\rm{\gamma max}})}\rangle$) is calculated as the sum of  $\langle{\sigma(E_{\rm{\gamma max}})}\rangle_{\rm{g}}$ and $\langle{\sigma(E_{\rm{\gamma max}})}\rangle_{\rm{m}}$  average cross-sections for population of the ground state and the isomeric state, respectively.     

When calculating the values of the bremsstrahlung flux-averaged cross-sections, it was assumed that all the radioisotopes were produced only as a result of photonuclear reactions on $^{181}\rm{Ta}$, because the $^{180\rm{m}}\rm{Ta}$ isomer content in the natural tantalum mix is negligibly small (0.012\%). The self-absorption of $\gamma$-radiation of the reaction products in the target was calculted with the GEANT4 code, and was taken into account in calculations by eq.~\ref{form1}.          

The bremsstrahlung flux monitoring by the  $^{100}\rm{Mo}(\gamma,n)^{99}\rm{Mo}$ reaction yield was performed by comparing the experimentally obtained average cross-section values with the computation data. To determine the experimental $\langle{\sigma(E_{\rm{\gamma max}})}\rangle_{\rm{exp}}$ values by eq.~\ref{form1}, we have used the activity $\triangle A$ for the $\gamma$-line of energy $E_{\gamma} = 739.50$~keV and absolute intensity $I_{\gamma} = 12.13\%$ (see Table~\ref{tab1}). The average cross-section $\langle{\sigma(E_{\rm{\gamma max}})}\rangle_{\rm{th}}$ values were computed by eq.~\ref{form2} with the cross-sections $\sigma(E)$ from the TALYS1.9 code. The obtained normalization factor  $k = \langle{\sigma(E_{\rm{\gamma max}})}\rangle_{\rm{th}} /\langle{\sigma(E_{\rm{\gamma max}})}\rangle_{\rm{exp}}$ represents the deviation of the GEANT4-computed bremsstrahlung flux from the real $\gamma$-flux incident on the target. The thus found $k$ values, which varied within $1.08 \div 1.15$, were used for normalization of the cross-sections for photoneutron reactions on the  $^{181}\rm{Ta}$ nucleus. The details of the monitoring procedure can be found in ref.~\cite{24,Nb}.

The Ta-converter and Al-absorber, used in the experiment, generates neutrons that can cause the reaction  $^{100}\rm{Mo}(n,2n)^{99}\rm{Mo}$. Calculations were made of the energy neutrons spectrum and the fraction of neutrons with energies above the threshold of this reaction. The contribution of the $^{100}\rm{Mo}(n,2n)^{99}\rm{Mo}$ reaction to the value of the induced activity of the $^{99}$Mo nucleus has been estimated and it has been shown that this contribution is negligible compared to the contribution of  $^{100}\rm{Mo}(\gamma,n)^{99}\rm{Mo}$.

The uncertainty of measurements of the average cross-sections $\langle{\sigma(E_{\rm{\gamma max}})}\rangle$, $\langle{\sigma(E_{\rm{\gamma max}})}\rangle_{\rm{g}}$ and $\langle{\sigma(E_{\rm{\gamma max}})}\rangle_{\rm{m}}$ was determined as a squared sum of statistical and systematic errors. The statistical error in the observed $\gamma$-activity is mainly associated with the statistics calculation for the total-absorption peak of the corresponding $\gamma$-line, which varies from 1 to 10\%. The error varies depending on the $\gamma$-line intensity and the background conditions of the spectrum measurement. The $\gamma$-line intensity depends on the registration efficiency, the half-life period and the absolute intensity $I_{\gamma}$. The background is mainly contributed by the Compton scattering of high-energy quanta. 

The systematic errors stem from the uncertainties in:
\begin{enumerate} 
  \item	irradiation time -- $0.25 \div 0.5\%$; 
  \item 	electron current -- 0.5\%;
  \item	$\gamma$-radiation detection efficiency $\sim 2 \div 3\%$, mainly due to the measuring error of $\gamma$-radiation sources;
  \item 	half-life period $T_{1/2}$ of reaction products and the absolute intensity of the analyzed $\gamma$-quanta $I_{\gamma}$  -- $1.0 \div 20\%$ (see Table~\ref{tab1});
 \item	normalization of the experimental data to the yield of the monitoring reaction  $^{100}\rm{Mo}(\gamma,n)^{99}\rm{Mo}$ -- 2.5\%;
 \item	the GEANT4 computational error for the bremsstrahlung spectra $\sim 1.5\%$.
\end{enumerate}

So, the statistical and systematic errors represent the variables and are different for different $^{181}\rm{Ta}(\gamma,\textit{x}n; \textit{x} = 1 \div 8)^{181-\textit{x}}\rm{Ta}$  reactions. The overall uncertainty of the experimental data is given in Figs.~\ref{fig3}(a-h) and \ref{fig5}(a).

 \section{RESULTS AND DISCUSSION}
 \label{RES AND DISC} 
\subsection{The total bremsstrahlung flux-averaged cross-sections  $\langle{\sigma(E_{\rm{\gamma max}})}\rangle$ for the reactions  $^{181}\rm{Ta}(\gamma,\textit{x}n; \textit{x} = 1 \div 8)^{181-\textit{x}}\rm{Ta}$}
 \label{subsec1}     

\begin{figure*}[]
  	\resizebox{0.9\textwidth}{!}{%
  \includegraphics{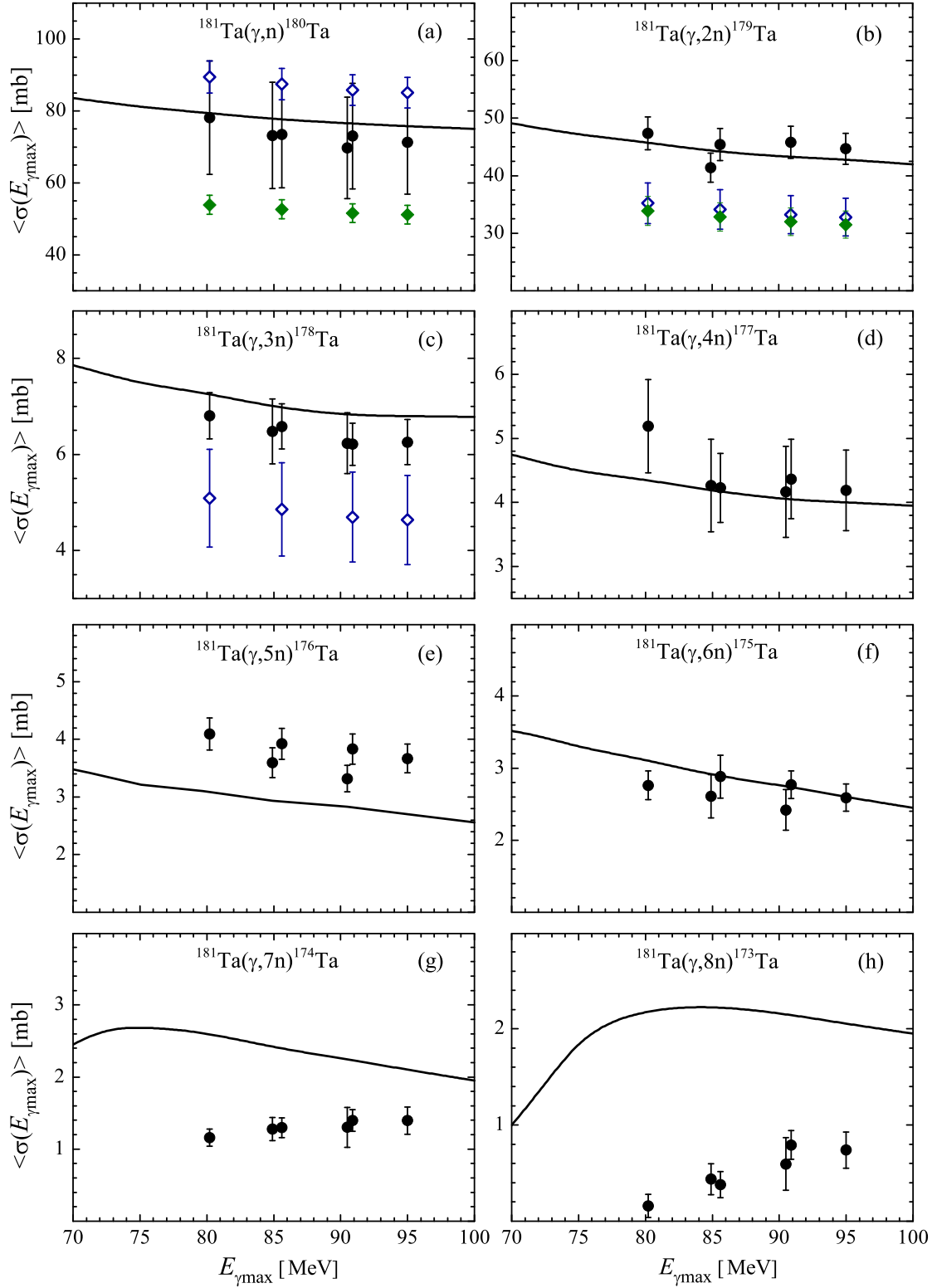}}
	\vspace{-2ex}
	\caption{Total bremsstrahlung flux-averaged cross-sections $\langle{\sigma(E_{\rm{\gamma max}})}\rangle$ for the  $^{181}\rm{Ta}(\gamma,\textit{x}n; \textit{x} = 1 \div 8)^{181-\textit{x}}\rm{Ta}$ reactions.
Solid circles show the present experimental data. The calculated average cross-sections are shown by: the curves – with the use of $\sigma(E)$ from the TALYS1.9 code; solid diamonds – with the use of the partial cross-section data from ref.~\cite{7}; empty diamonds – partial cross-sections from ref.~\cite{8}.}
	\label{fig3}
\end{figure*}

The obtained experimental data on the total average cross-sections  $\langle{\sigma(E_{\rm{\gamma max}})}\rangle$ for the $^{181}\rm{Ta}(\gamma,\textit{x}n; \textit{x} = 1 \div 8)^{181-\textit{x}}\rm{Ta}$ reactions in the end-point energy range of bremsstrahlung $\gamma$-quanta $E_{\rm{\gamma max}} = 80 \div 95$~MeV are presented in Fig.~\ref{fig3}. The results of calculations by eq.~\ref{form2} with the use of cross-sections $\sigma(E)$ computed with the TALYS 1.9 code are shown in the same figure.

In the $^{181}\rm{Ta}(\gamma,n)^{180}\rm{Ta}$ reaction case, one can measure only $\langle{ \sigma(E_{\rm{\gamma max}}) }\rangle_{\rm{g}}$ for occupation of the ground state of the $^{180\rm{g}}\rm{Ta}$ nucleus.  The total average cross-section  $\langle{\sigma(E_{\rm{\gamma max}})}\rangle$  for this reaction has been calculated with the use of the theoretical value of  $\langle{\sigma(E_{\rm{\gamma max}})}\rangle_{\rm{m}}$ for occupation of the isomeric state of the nucleus-product. According to the TALYS1.9 computations, the ratio of the cross-sections  $\langle{\sigma(E_{\rm{\gamma max}})}\rangle_{\rm{m}}$ and  $\langle{\sigma(E_{\rm{\gamma max}})}\rangle$  has made 0.083 for our experimental conditions. This value is in satisfactory agreement with the value 0.07 from ref.~\cite{16}. In the  $^{181}\rm{Ta}(\gamma,3n)^{178}\rm{Ta}$ case, Fig.~\ref{fig3}(c) shows the values of the total cross-section $\langle{\sigma(E_{\rm{\gamma max}})}\rangle$, while its components $\langle{\sigma(E_{\rm{\gamma max}})}\rangle_{\rm{g}}$ and $\langle{ \sigma(E_{\rm{\gamma max}}) }\rangle_{\rm{m}}$  are presented in Fig.~\ref{fig5}(a).

As is evident from Fig.~\ref{fig3}, the obtained  $\langle{ \sigma(E_{\rm{\gamma max}}) }\rangle$ values are in satisfactory agreement (within the experimental uncertainties) with the calculations for the  $^{181}\rm{Ta}(\gamma,\textit{x}n)^{181-\textit{x}}\rm{Ta}$ reactions with the escape of 1 to 6 neutrons. With the increase in the number of escaping neutrons up to 7 and 8 in the exit channel of the reactions, a significant difference is observed between the experiment and the calculation. Note also the tendency to satisfactory agreement between the experimental and calculated data on the total cross-sections for photoneutron reactions on $^{181}\rm{Ta}$, in which the nuclei-products are produced with positive parity $\pi$ in the ground state.
   
So far, no data can be found in the literature on the experimental average cross-sections for tantalum photodisintegration. However, using eq.~\ref{form2} and the data on partial cross-sections $\sigma(E)$ for photonuclear reactions on the $^{181}\rm{Ta}$ nucleus with escape of up to 3 neutrons (refs.~\cite{7,8}), it appears possible to calculate the values of total average cross-sections  $\langle{ \sigma(E_{\rm{\gamma max}}) }\rangle$ for the bremsstrahlung flux consistent with the conditions of the given experiment. The calculated in this way values for all three reactions exhibit significant deviations from both our experimental results and the calculations performed with the use of the TALYS1.9 code (see Figs.~\ref{fig3}(a-c)).

\begin{figure}[t]
  	\resizebox{0.5\textwidth}{!}{%
  \includegraphics{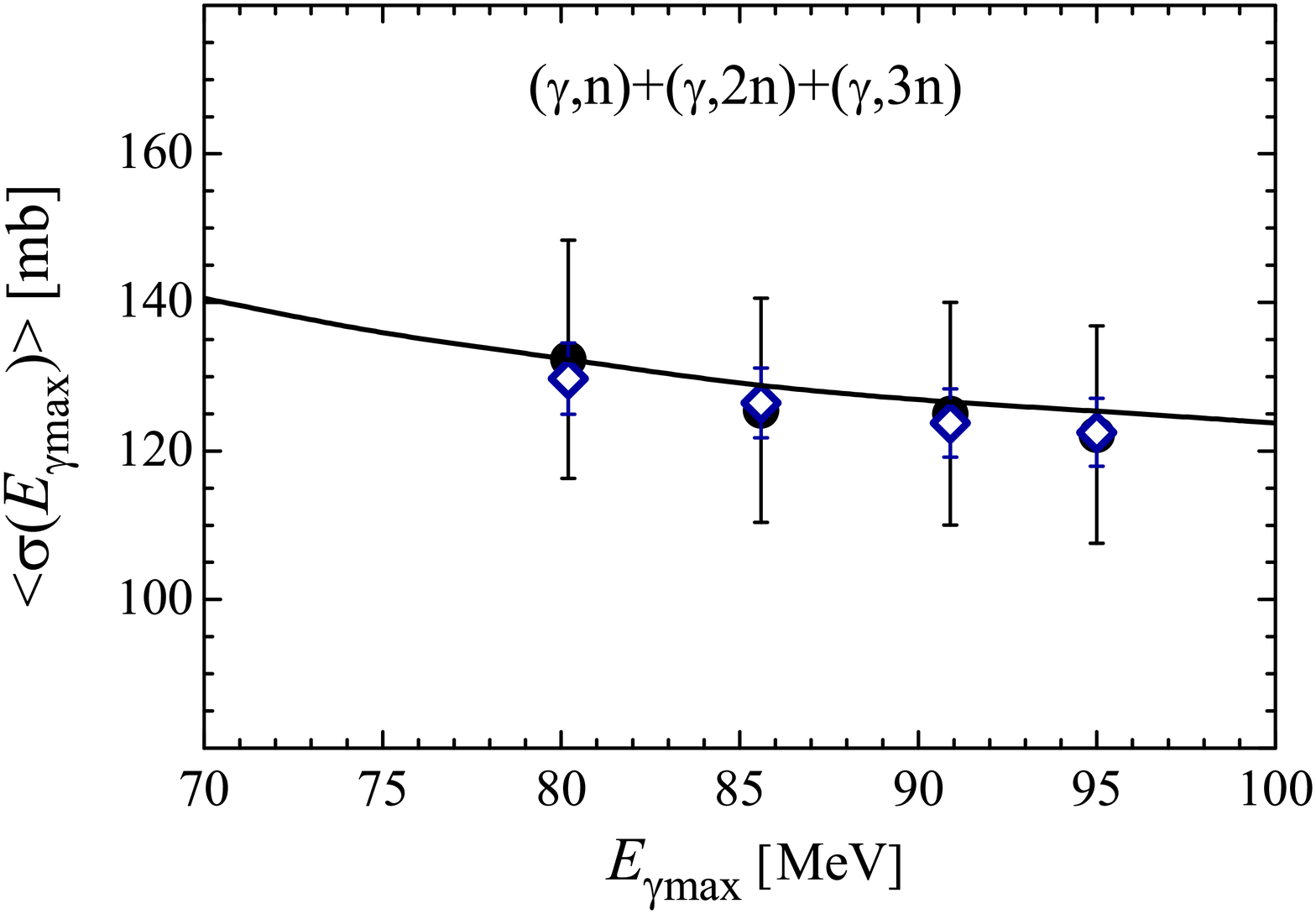}}
	\caption{The sum of three total average cross-sections for the photoneutron reactions on $^{181}\rm{Ta}$: $(\gamma,\rm{n})$+$(\gamma,2\rm{n})$+$(\gamma,3\rm{n})$. Solid circles -- the present data, diamonds -- with the use of partial cross-sections from ref.~\cite{8}; the curve shows the computation with $\sigma(E)$ from the TALYS1.9 code.}
	\label{fig4}
\end{figure}

The analysis of the experimental data from refs.~\cite{7,8} shows their difference for the $(\gamma,\rm{n})$ reaction, viz., at the maximum of the cross-section the difference reaches $\sim~25\%$. This may be attributed to complexity of neutron separation from the pairs of the $(\gamma,\rm{n})$ and $(\gamma,2\rm{n})$ reactions in the method of direct neutron registration. Another reason of the difference, according to ref.~\cite{16}, may be probably due to incorrect accounting of the bremsstrahlung part of the quasimonochromatic photon spectrum. On this basis it can be assumed that the sum of total average cross-sections $\langle{ \sigma(E_{\rm{\gamma max}}) }\rangle$ for the reactions $(\gamma,\rm{n})$, $(\gamma,2\rm{n})$ and $(\gamma,3\rm{n})$, which were obtained from the data of ref.~\cite{8}, 
 should be close to a similar sum obtained from our data. In the energy range $E_{\rm{\gamma max}} = 80 \div 95$~MeV, 
the sums of three cross-sections $(\gamma,\rm{n})+(\gamma,2\rm{n})+(\gamma,3\rm{n})$, calculated with the data from ref.~\cite{8}, smoothly decrease from $130 \pm 6$ mb down to $123 \pm 6$~mb with energy increase. With the data found in the present work, the sums of the three total average cross-sections vary from $ 132 \pm 16$~mb to $122  \pm 15$~mb. It is apparent that there is close agreement with the values of the summed total cross-sections (see Fig.~\ref{fig4}). The similar summed cross-section computed with $\sigma(E)$ from the TALYS1.9 code is in agreement with the results based on the experimental data. 

  \subsection{Isomeric ratio $d(E_{\rm{\gamma max}})$ of average cross-sections for the products of the  $^{181}\rm{Ta}(\gamma,3n)^{178g,m}\rm{Ta}$ reactions}
  \label{subsec2}
  
\begin{figure}[]
  	\resizebox{0.5\textwidth}{!}{%
  \includegraphics{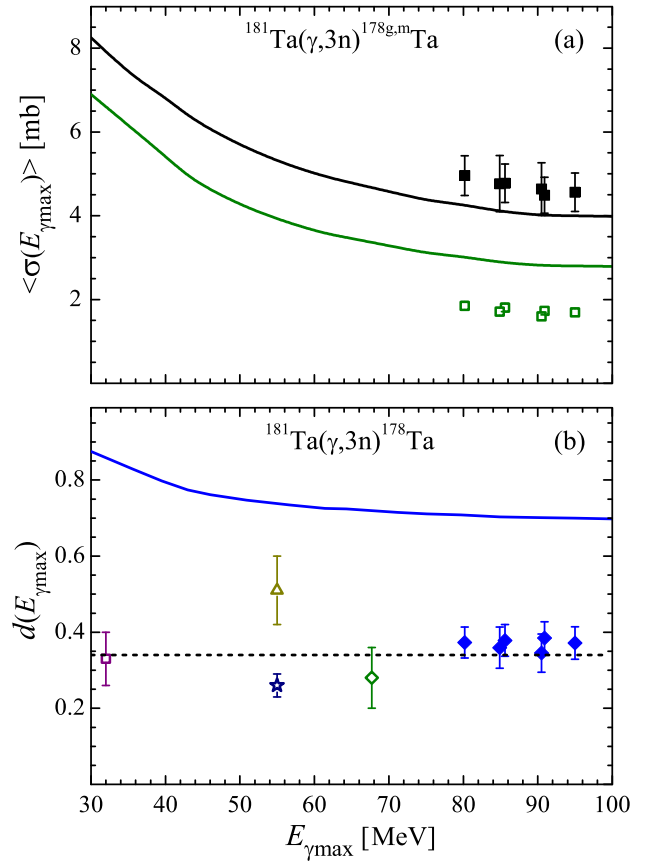}}
	\caption{ Reactions  $^{181}\rm{Ta}(\gamma,3n)^{178g,m}\rm{Ta}$.\\
  (a) The bremsstrahlung flux-averaged cross-sections for occupation of the ground  $\langle{ \sigma(E_{\rm{\gamma max}}) }\rangle_{\rm{g}}$ and isomeric $\langle{ \sigma(E_{\rm{\gamma max}}) }\rangle_{\rm{m}}$ states of the nucleus-product $^{178}\rm{Ta}$. Full squares represent $\langle{ \sigma(E_{\rm{\gamma max}}) }\rangle_{\rm{g}}$, empty squares -- $\langle{ \sigma(E_{\rm{\gamma max}}) }\rangle_{\rm{m}}$  (experimental error is smaller than the symbol size)), the lines show the corresponding computations with TALYS1.9 code. \\
(b) Isomeric ratios $d(E_{\rm{\gamma max}})$ of average cross-sections. Full diamonds show the present results, star – ref.~\cite{15}, empty diamond  – ref.~\cite{16}, square – ref.~\cite{21}, triangle – ref.~\cite{22}. Solid line represents the calculation, dashed line – experimental data fitting. }
	\label{fig5}
\end{figure}

 The values of average cross-sections have been obtained experimentally for occupation of both the ground  $\langle{ \sigma(E_{\rm{\gamma max}}) }\rangle_{\rm{g}}$ and isomeric $\langle{ \sigma(E_{\rm{\gamma max}}) }\rangle_{\rm{m}}$ states of the nucleus-product in the $^{181}\rm{Ta}(\gamma,3n)$ reaction (see Fig.~\ref{fig5}(a)). Comparison with the calculation by eq.~\ref{form2} for the  $^{181}\rm{Ta}(\gamma,3n)^{178g}\rm{Ta}$ reaction shows that the experimental $\langle{ \sigma(E_{\rm{\gamma max}}) }\rangle_{\rm{g}}$ values slightly exceed the calculated data. On the average, this is within the magnitude of the experimental accuracy. At the same time, for $\langle{ \sigma(E_{\rm{\gamma max}}) }\rangle_{\rm{m}}$ the experiment gives the value substantially lower than the theoretical evaluation throughout the energy range under study. However, as is seen from Fig.~\ref{fig3}(c), the total average cross-section values agree within the experimental accuracy with the calculation. 

On the basis of the obtained $\langle{ \sigma(E_{\rm{\gamma max}}) }\rangle_{\rm{m}}$ and  $\langle{ \sigma(E_{\rm{\gamma max}}) }\rangle_{\rm{g}}$ values for the $^{181}\rm{Ta}(\gamma,3n)^{178g,m}\rm{Ta}$ reactions, we have determined the isomeric ratio values for average cross-sections $d(E_{\rm{\gamma max}}) = \langle{ \sigma(E_{\rm{\gamma max}}) }\rangle_{\rm{m}} / \langle{ \sigma(E_{\rm{\gamma max}}) }\rangle_{\rm{g}}$ (see Fig.~\ref{fig5}(b)). Within the limits of the experimental uncertainty, the $d(E_{\rm{\gamma max}})$ values are constant and equal to $0.37 \pm 0.02$ over the investigated energy range $E_{\rm{\gamma max}} = 80 \div 95$~MeV. This estimated value points to a significant (by a factor of $\sim 2.5$) suppression of the reaction with occupation of the isomeric state of the nucleus $^{178\rm{m}}\rm{Ta}$  ($J^{\pi} = (7)^{–}$) relative to the ground-state occupation of  $^{178\rm{g}}\rm{Ta}$ ($J^{\pi} = (1)^+$).

For comparison with theory, the $d(E_{\rm{\gamma max}})$ values were computed with the use of the TALYS1.9 code (see Fig.~\ref{fig5}(b)). The computed $d(E_{\rm{\gamma max}})$ value has appeared to be nearly two-fold higher than the experimental one. Most likely, the main reason for this difference, as it follows from Fig.~\ref{fig5}(a), lies in the TALYS1.9 overestimation of the $\langle{ \sigma(E_{\rm{\gamma max}}) }\rangle_{\rm{m}}$ cross-section values for occupation of the isomeric state of the nucleus-product. 

The earlier obtained values of the isomeric ratio $d(E_{\rm{\gamma max}})$ of the $^{181}\rm{Ta}(\gamma,3n)^{178g,m}\rm{Ta}$ reaction yields can be found in refs.~\cite{15,16,21,22}. Note that in ref.~\cite{21} the  $d(E_{\rm{\gamma max}})$ value was obtained in the range $E_{\rm{\gamma max}} = 24 \div 32$~MeV, and according to the authors’ conclusions, the value 
$d(E_{\rm{\gamma max}}) = 0.33 \pm 0.07$ is constant in the energy region under study (in Fig.~\ref{fig5}(b) the $d(E_{\rm{\gamma max}})$ was given only at $E_{\rm{\gamma max}} = 32$~MeV). From the data of paper \cite{16}, we have estimated the $d(E_{\rm{\gamma max}})$ value to be $0.28  \pm  0.08$ at $E_{\rm{\gamma max}} = 67.7$~MeV. The experimental $d(E_{\rm{\gamma max}})$ values from refs.~\cite{15,16,21,22} are in agreement with our present data for the range $E_{\rm{\gamma max}} = 80 \div 95$~MeV, and also, similarly to the present results, are systematically below the calculated values.  

Figure~\ref{fig5}(b) shows that the calculated value of the isomeric ratio $d(E_{\rm{\gamma max}})$ increases with a decrease in the end-point bremsstrahlung energies  $E_{\rm{\gamma max}}$, however the available experimental data do not support this dynamics. The experimental $d(E_{\rm{\gamma max}})$ values in the energy range $E_{\rm{\gamma max}} = 24 \div 95$~MeV are grouped around the constant value equal to $0.34 \pm 0.02$. 

\section{Conclusions}
\label{Concl}
The present work has been concerned with investigation of the $^{181}\rm{Ta}$ photoneutron reactions with escape of up to 8 neutrons at end-point bremsstrahlung energies $E_{\rm{\gamma max}} = 80 \div 95$~MeV with the use of the residual $\gamma$--activity method. The obtained total bremsstrahlung flux-averaged cross-section $\langle{ \sigma(E_{\rm{\gamma max}}) }\rangle$ values have been compared with the computations using $\sigma(E)$ from the TALYS1.9 code with the default options. A satisfactory agreement has been demonstrated for the $^{181}\rm{Ta}(\gamma,\textit{x}n)^{181-\textit{x}}\rm{Ta}$ reactions with escape of 1 to 6 neutrons. With increase in the number of neutrons up to 7 and 8 in the exit channel of the reactions, a significant difference is observed between the experiment and the calculation. The tendency has been traced to satisfactory agreement between the experimental and calculated data on the bremsstrahlung flux-averaged cross-sections for photoneutron reactions on $^{181}\rm{Ta}$, in which the nuclei-products are produced with positive parity $\pi$ in the ground state.

The total bremsstrahlung flux-averaged cross-sections $\langle{ \sigma(E_{\rm{\gamma max}}) }\rangle$ measured in the present work for the reactions $(\gamma,\rm{n})$, $(\gamma,2\rm{n})$ and $(\gamma,3\rm{n})$ are different from the data obtained on the basis of partial cross-sections from refs.~\cite{7,8}. However, the value resulting from the summation of our total bremsstrahlung flux-averaged cross-sections  ($(\gamma,\rm{n})$+$(\gamma,2\rm{n})$+$(\gamma,3\rm{n})$) coincides with a similar sum from ref.~\cite{8}.

Isomeric ratios of average cross-sections $d(E_{\rm{\gamma max}})$ for the $^{181}\rm{Ta}(\gamma,3n)^{178g,m}\rm{Ta}$ reaction products have been obtained in the energy range $E_{\rm{\gamma max}} = 80 \div 95$~MeV. The measured values point to a significant (by a factor of $\sim~2.5$) suppression of the occupation of the isomeric state of the nucleus $^{178\rm{m}}\rm{Ta}$ in relation to the ground-state occupation of $^{178\rm{g}}\rm{Ta}$. The comparison between the experimental results and the TALYS1.9 code-based computed values has shown a two-fold difference between the data. The present $d(E_{\rm{\gamma max}})$ values are in agreement with the literature data \cite{15,16,21,22}. All the available experimental $d(E_{\rm{\gamma max}})$ values in the range  $E_{\rm{\gamma max}} = 24 \div 95$~MeV are grouped around the constant value $0.34 \pm 0.02$.

The experimental data on the bremsstrahlung flux-averaged cross-sections $\langle{ \sigma(E_{\rm{\gamma max}}) }\rangle$,  $\langle{ \sigma(E_{\rm{\gamma max}}) }\rangle_{\rm{g}}$,  $\langle{ \sigma(E_{\rm{\gamma max}}) }\rangle_{\rm{m}}$ and the isomeric ratios of the average cross-sections $d(E_{\rm{\gamma max}})$ at $E_{\rm{\gamma max}}$ energies ranging from 80 to 95 MeV have been obtained for the first time. 
 
%
%

\end{document}